\RequirePackage{etoolbox}
\let\originalshowhyphens\showhyphens

\documentclass[sigconf]{acmart}
\acmConference[EASE 2026]{The 30th International Conference on Evaluation and Assessment in Software Engineering}{9–12 June, 2026}{Glasgow, Scotland, United Kingdom}
\acmBooktitle{The 30th International Conference on Evaluation and Assessment in Software Engineering (EASE '26), 9–12 June, 2026, Glasgow, Scotland, United Kingdom}
\acmDOI{}
\acmISBN{}

\let\showhyphens\originalshowhyphens
\usepackage{hyperref}
\usepackage{graphicx}
\usepackage{booktabs}
\usepackage{tcolorbox}
\usepackage{multirow}
\usepackage{siunitx}
\usepackage{enumitem}
\usepackage[table]{xcolor}

\makeatletter

\newcolumntype{L}[1]{>{\raggedright\let\newline\\\arraybackslash\hspace{0pt}}m{#1}}
\newcolumntype{C}[1]{>{\centering\let\newline\\\arraybackslash\hspace{0pt}}m{#1}}
\newcolumntype{R}[1]{>{\raggedleft\let\newline\\\arraybackslash\hspace{0pt}}m{#1}}

\newcommand{\etal}{et~al.}
\newcommand{\up}{$\nearrow$}
\newcommand{\dn}{$\searrow$}

\usepackage{framed}
\definecolor{shadecolor}{gray}{0.95}
\newenvironment{rqbox}{
  \par\noindent
  \setlength{\FrameRule}{0.4pt}
  \setlength{\FrameSep}{6pt}
  \begin{shaded*}\noindent
}{
  \end{shaded*}
}

\makeatother

\begin{document}

\title[The Promise and Reality of Continuous Integration Caching]{The Promise and Reality of Continuous Integration Caching: An~Empirical Study of Travis~CI Builds}

\author{Taher A. Ghaleb}
\orcid{0000-0001-9336-7298}
\affiliation{
  \department{Department of Computer Science} 
  \institution{Trent University}
  \city{Peterborough}
  \state{Ontario}
  \country{Canada}}
\email{taherghaleb@trentu.ca}

\author{Daniel Alencar da Costa}
\orcid{0000-0003-4525-3266}
\affiliation{
  \department{Information Science Department} 
  \institution{University of Otago}
  \city{Dunedin}
  \country{New Zealand}}
\email{danielcalencar@otago.ac.nz}

\author{Ying Zou}
\affiliation{
  \department{Department of Electrical and Computer Engineering} 
  \institution{Queen's University}
  \city{Kingston}
  \state{Ontario}
  \country{Canada}}
\email{ying.zou@queensu.ca}

\begin{abstract}
Continuous Integration (CI) provides early feedback by automatically building software, but long build durations can hinder developer productivity. CI services use caching to speed up builds by reusing infrequently changing artifacts, yet little is known about how caching is adopted in practice and what challenges it entails. In this paper, we conduct a large-scale empirical study of CI caching in Travis~CI, analyzing 513,384 builds from 1,279 GitHub projects. We find that only 30\% of projects adopt CI caching, and early adopters are typically more mature, with more dependencies, commits, and longer CI lifespans. To understand non-adoption, we submit pull requests enabling caching in non-adopting projects, and nearly half are accepted or merged. Developer feedback indicates that non- or late adoption mainly results from limited awareness of CI caching support. We further study cache maintenance and identify five common activities, performed by 24\% of cache-enabled projects. While one-third of projects see substantial build-time reductions, cache uploads occur in 97\% of builds, and 27\% of projects contain stale cached artifacts. An analysis of reported caching issues shows developers mainly struggle with corrupted or outdated caches and request broader caching features. Overall, CI caching does not benefit all projects, requires ongoing maintenance, and is more complex in practice than many developers expect.
\end{abstract}



\maketitle

\section{Introduction}
Continuous Integration (CI) enables developers to integrate code changes into remote repositories, automatically generate builds, and detect errors early~\cite{fowler2006continuous}. However, CI builds can be slow~\cite{beller2017oops,ghaleb2019duration,ghaleb2022interplay,aidasso2025build}, wasting resources and hindering other development activities. CI services provide caching for infrequently changing artifacts~\cite{travis_caching}. As reported by prior work, caching can speed up builds~\cite{esfahani2016cloudbuild,ghaleb2019duration,gallaba2020accelerating}. However, caching is not enabled by default; developers must configure which artifacts to cache. Yet, developers may (a) lack awareness of which artifacts change less frequently, (b) misuse caching configurations, and (c) invest time/effort maintaining the cache throughout the project's lifetime. Moreover, prior research has not established whether caching is always beneficial. As the \textsc{Travis~CI} team states: ``\textit{There is pros and cons of using the cache. It is up to you [developers] to decide whether you want to use it}".\footnote{\href{https://github.com/travis-ci/travis-ci/issues/6396\#issuecomment-240926751}{travis-ci/travis-ci/issues/6396\#issuecomment-240926751}} Although CI caching can speed up builds, its associated challenges (e.g., maintenance effort) and issues (e.g., unexpected failures) remain underexplored. There is little empirical guidance on how to adopt caching carefully, how well it is adopted and maintained in open source projects, and whether it benefits every project. Studying current practices, challenges, and issues of CI caching can help developers set realistic expectations and avoid unnecessary overhead.

In this paper, we empirically study the adoption of CI caching in {\sc Travis~CI}, a cloud-based CI service. We analyze $513,384$ builds from $1,279$ {\sc GitHub} projects using {\sc Travis~CI}. We investigate the adoption rate of CI caching to understand why some projects adopt it early or proactively, and why others do not adopt or delay adoption.
We group the studied projects into five categories based on caching adoption and model differences among these categories using logistic regression.
To gain further insights on caching adoption, we submit 85 pull requests to enable CI \textit{cache} in projects that had never used caching, and we comment on pull requests/commits where CI caching was adopted proactively or belatedly, asking for justifications.
To understand potential overhead, we also analyze cache maintenance activities, the impact of caching on CI builds, and commonly reported CI caching issues.

\vspace{3pt}
\noindent\textbf{Research Questions ({\textit{RQ\textsubscript{s}}}).}
~\noindent 
We study current CI caching practices and challenges across five research questions (RQs):

    \vspace{2pt}
    \noindent\textbf{{\textit{RQ\textsubscript{1}}}: \emph{What urges software projects to adopt CI caching?} }
        CI caching can speed up builds, yet 70\% of projects in our dataset do not use it, mainly due to lack of awareness of CI caching support. As a result, developers of these projects incur opportunity costs by disregarding the benefits of caching.
                
    \vspace{2pt}
    \noindent\textbf{{\textit{RQ\textsubscript{2}}}: \emph{How do developers maintain CI caching?}}
        Even with caching enabled, developers must regularly update caching configurations to match code changes. However, maintenance practices for CI caching are not well understood. By mining changes to CI caching configurations in our projects, we identify five common maintenance activity patterns. Still, most projects (76\%) perform no maintenance after adoption. Developers should regularly maintain cached artifacts to preserve the benefits of CI caching.
            
    \vspace{2pt}
    \noindent\textbf{{\textit{RQ\textsubscript{3}}}: \emph{To what extent does CI caching reduce the build duration?}}
         Prior research has paid little attention to whether CI caching consistently reduces build duration. We model the change in build time before and after adopting CI caching to assess its impact. Two-thirds of projects see no reduction, and 80\% of those do not maintain CI caching regularly. Developers should not treat CI caching as a “\textit{one size fits all}” solution.

    \vspace{2pt}
    \noindent\textbf{{\textit{RQ\textsubscript{4}}}: \emph{How much overhead does CI caching introduce to builds?}}
        CI caching adds steps such as downloading and uploading caches, but their impact on build time is unclear. CI log analysis shows that cache uploads occur in 97\% of builds and take six times longer than downloads. We find that frequent cache uploads stem from caching rapidly changing artifacts, such as installation logs. CI services should assist developers in monitoring CI logs for abnormal cache uploads and exclude unnecessary artifacts from the CI cache.

    \vspace{2pt}
    \noindent\textbf{{\textit{RQ\textsubscript{5}}}: \emph{What issues do developers encounter with CI caching?}}
        Adopting CI caching can introduce problems such as occasional build failures. Identifying these problems helps developers avoid them and decide whether to adopt CI caching or not. In this RQ, we analyze caching-related issues reported to the {\sc Travis~CI} team. Our results show common issues with cache storage, where artifacts become corrupted or outdated. The CI team recommends regularly clearing the CI cache, which resolves most caching issues.
        
\vspace{3pt}
\noindent\textbf{In summary, this paper makes the following contributions:}
\begin{itemize}[leftmargin=18.5pt]
    \vspace{-1pt}
    \item An empirical study showing limited caching adoption in {\sc Travis~CI} and its association with project maturity.
    \item An intervention study showing that CI caching enabled via pull requests is accepted by nearly half of developers.
    \item A process-mining analysis revealing the ongoing maintenance required for CI caching.
    \item An impact analysis showing that adopting CI caching does not consistently reduce build duration.
    \item An analysis of issue reports identifying cache corruption and limited tool support as major challenges.
\end{itemize}

\vspace{3pt}
\noindent\textbf{Paper organization.}
The rest of this paper is organized as follows.
Section~\ref{background} presents background about CI.
Section~\ref{Experimental_Setup} describes the experimental setup of our study.
Section~\ref{Experimental_Results} presents the results and findings of our studied RQs.
Section~\ref{Discussion} discusses the implications of our findings.
Section~\ref{Threats_to_Validity} presents threats to the validity of our results.
Section~\ref{Related_Work} presents the related literature on CI builds.
Finally, Section~\ref{Conclusion} concludes the paper and outlines possible future work.

\vspace{-2pt}
\section{Background}
\label{background}

\noindent\textbf{Continuous Integration (CI).}
CI automates software builds and provides early feedback on code changes~\cite{fowler2006continuous}. Builds are typically triggered by commits or pull requests, fetching the source code, installing dependencies, building the software, and running tests. CI builds pass if all steps succeed, otherwise they fail.

\vspace{2pt}
\noindent\textbf{{T{\scriptsize RAVIS} CI}.}
{\sc Travis~CI}\footnote{\url{https://travis-ci.com}} is a widely used cloud-based CI service~\cite{GitHubCI}. Its build lifecycle consists of two main phases, \texttt{install} and \texttt{script}, plus an optional \texttt{deploy} phase. During \texttt{install}, the repository is cloned and dependencies installed; during \texttt{script}, code is built and tests run; \texttt{deploy} packages and releases software. Developers customize builds via \texttt{.travis.yml}, including machine specifications, timeouts, and CI features such as caching.

\vspace{2pt}
\noindent\textbf{CI Caching.}
On December 5, 2013, {\sc Travis~CI} introduced caching for private repositories~\cite{travis_cache_2013}. Support for public repositories followed on December 17, 2014, with the introduction of container-based infrastructure~\cite{travis_cache_2014}. Finally, caching became available across all infrastructures on May 3, 2016~\cite{travis_cache_2016}.
Developers enable caching with the \texttt{cache:} directive, typically for dependencies~\cite{travis_caching}, and use \texttt{before\_cache:} to delete temporary files, such as logs. Cached artifacts are uploaded after a build, reused in later builds, and only changed files are re-uploaded. Builds proceed even if cache operations (downloads or uploads) fail, and caches can be invalidated via the API or on the web. Each CI job has its own cache.
Other popular CI services, such as {\sc GitHub~Actions} and {\sc CircleCI}, provide similar build customization and caching mechanisms, where developers can define cache contents, control storage/retrieval, and clear outdated caches. Given the shared principles, we expect our findings on {\sc Travis~CI} caching to have broader relevance, offering guidance for optimizing caching strategies in other CI environments.

\section{Experimental Setup}
\label{Experimental_Setup}
This section describes the experimental setup of our study and how we collect and prepare the data for our RQs.

\subsection{\large Data Collection}
\label{Data_Collection}
\begin{figure*}
  \centering
  \resizebox{1.005\linewidth}{!}{
  \includegraphics{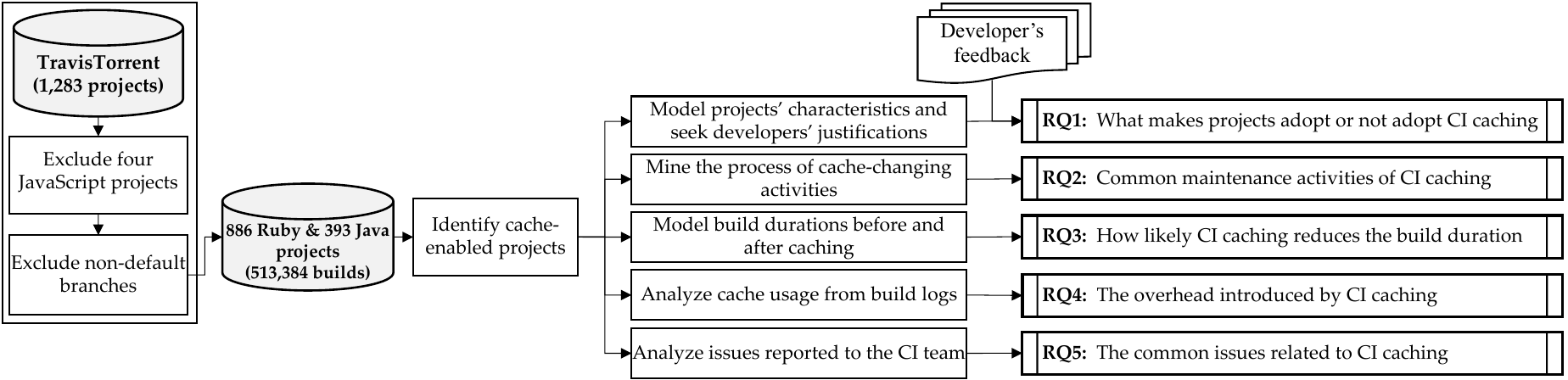}
  }
  \vspace{-15pt}
  \caption{Overview of our study}
  \label{ProcessModel}
  \vspace{-7pt}
\end{figure*}

Figure~\ref{ProcessModel} shows an overview of our study, which uses data from {\sc TravisTorrent}~\cite{beller2017travistorrent}, a common dataset for CI build research~\cite{beller2017oops,ni2017cost,luo2017factors}. {\sc TravisTorrent}~includes builds from $1,283$ projects: $886$ Ruby, $393$ Java, and $4$ JavaScript. 
We exclude the four JavaScript projects, since they are not a representative sample of that language.
Although the last build in our dataset was triggered on $Aug$ $31$, $2016$, we further verified in 2021 that all projects remain active and maintained by examining their commit history and issue activity. We also interacted with developers from both caching adopters and non-adopters by submitting pull requests to non-adopters and commenting on commits where caching was enabled, receiving responses from over half of the contacted projects.
Each {\sc TravisTorrent} project includes builds from multiple branches. We consider only the default (e.g., \textit{main} or \textit{master}) branches, since they are more active and contain more builds. We identify each repository's default branch using the {\sc GitHub~API}.\footnote{\url{https://docs.github.com}} This yields $513,384$ builds. For each build, we collect the CI build job logs using the {\sc Travis API}.\footnote{\url{https://docs.travis-ci.com/api}}

\subsection{\large Data Processing}
\label{Metrics}
This section explains how we process data from the $1,279$ projects.

\subsubsection{\textbf{Computing project-level metrics}}
\vspace{-1pt}
\label{context_factors}
Our study conducts project-level analyses of CI caching adoption. Specifically, we:

\begin{itemize}[leftmargin=18.5pt]
    \item detect whether a project adopts CI caching by analyzing its \texttt{.travis.yml} configuration file;
    \item determine when a project adopts CI caching by locating the first commit that enables it; and
    \item identify characteristics of projects that adopt or do not adopt CI caching by analyzing each project's last build before adoption.
\end{itemize}

We examine how project characteristics are associated with CI caching adoption. For each characteristic, we compute $13$ project-level metrics (see Table~\ref{tab:project_level_metrics}) from builds triggered before caching adoption. Our data and scripts are publicly available in our replication package~\cite{our_replication_package}.

\begin{itemize}[leftmargin=18.5pt]
    \item \textit{Programming Language:} We analyze whether CI caching adoption is associated with a project's programming language, since caching needs can vary by language.

    \item \textit{Code Maturity:} We investigate whether code maturity (e.g., SLOC, test density, number of dependencies) is related to CI caching adoption.

    \item \textit{Development Activity:} We study whether more active projects (e.g., with more frequent commits) adopt CI caching differently from less active projects.

    \item \textit{CI Activity:} We analyze whether CI activity (e.g., build configurations and durations) is associated with CI caching adoption.
\end{itemize}

\begin{table*}
    \renewcommand{\arraystretch}{1}
	\centering
	\caption{Description of project-level metrics. All metrics are computed before a project adopts CI caching}
    \vspace{-8pt}
	\resizebox{0.9\textwidth}{!}{
	\begin{tabular}{lll}
	\toprule
    \textbf{Project characteristic} & \textbf{Project-level metric}    & \textbf{Description}\\
 	\midrule
	\multirow{1}{*}{Programming Language} 
                & Language                       & The {\sc GitHub} dominant programming language of a project\\[2pt]\cline{2-3}
 	
	\multirow{1}{*}{Code Maturity} 
                & Size (SLOC)                    & Number of source lines of code of a project\\
                & Dependencies                   & Number of dependencies (i.e., libraries, packages, or gems) a project relies on\\
			& Test density                   & Median number of test cases per $1,000$ SLOC of a project\\[2pt]\cline{2-3}
	
    \multirow{1}{*}{Development Activity}
                & \# of commits per lifetime     & Ratio of the number of commits per project age\\
                & Growth rate                    & Ratio of the relative increase/decrease (i.e., delta) in the lines of code\\
			& \# of developers               & Number of unique developers contributed to a project\\\cline{2-3}

    \multirow{1}{*}{CI Activity}
                & CI lifespan                    & Time (in terms of days) since a project adopted {\sc Travis~CI}\\
                & \# of builds                   & Number of CI builds triggered by a project\\
                & Building frequency             & Frequency (in terms of days) of triggering builds in a project\\
			& Configuration ratio            & Ratio of commits that change the build configuration file (\texttt{.travis.yml} per project lifetime\\
			& Build environments        & Number of unique integration environments used as build jobs in a project\\
			& Build duration                 & Median build duration of a project\\
 	\bottomrule
    \end{tabular}
    }
  \label{tab:project_level_metrics}
  \vspace{-6pt}
\end{table*}

\vspace{-1pt}
\subsubsection{\textbf{Performing correlation and redundancy analyses}}
\label{correlation_and_redundancy_analyses}
\vspace{-1pt}
We use project-level metrics as independent variables in our logistic regression models (see Section~\ref{RQ1}:\textit{RQ\textsubscript{1}}).
Following Harrell's regression modeling guidelines~\cite{harrell2001regression}, we first remove highly correlated variables, which can adversely affect regression models~\cite{domingos2012few}. We apply Spearman rank $\rho$ clustering~\cite{sarle1990varclus} (via the \texttt{varclus} function in the \texttt{rms} $R$ package~\cite{harrell_rms}) to identify highly correlated variables. For each pair with $|\rho| > 0.7$, we retain one variable and discard the other.
By the principle of parsimony, simple variables are preferred~\cite{vandekerckhove2015model}.
 Since all variables are similarly simple to compute, we keep those more informative about development and build processes. 
For instance, we exclude `\textit{\# of builds}', which is highly correlated with `\textit{\# of commits per lifetime}'.
We then perform redundancy analysis on the remaining variables, as redundant predictors can also distort regression models~\cite{harrell2001regression}. Using the \texttt{redun} function from the \texttt{rms} $R$ package, we search for variables that can be modeled by others with $R^2 \geq 0.9$ and find no redundant variables.

\vspace{-2pt}
\section{Empirical Analysis and Results}
\vspace{-2pt}
\label{Experimental_Results}
This section discusses RQ motivation, approaches, and findings.

\vspace{-4pt}
\subsection{\large RQ\textsubscript{1}: What urges software projects to adopt CI caching?}
\label{RQ1}

\subsubsection{\textbf{Motivation.}}
CI caching lets builds download software artifacts from the CI server instead of reinstalling them from original sources. Despite its potential speedup, some projects do not adopt CI caching. In this RQ, we (i) measure the proportion of projects that adopt CI caching and (ii) analyze factors associated with adopting, not adopting, or delaying its adoption.

\subsubsection{\textbf{Approach.}}
\begin{figure}
  \centering
  \resizebox{0.699\linewidth}{!}{
  \includegraphics{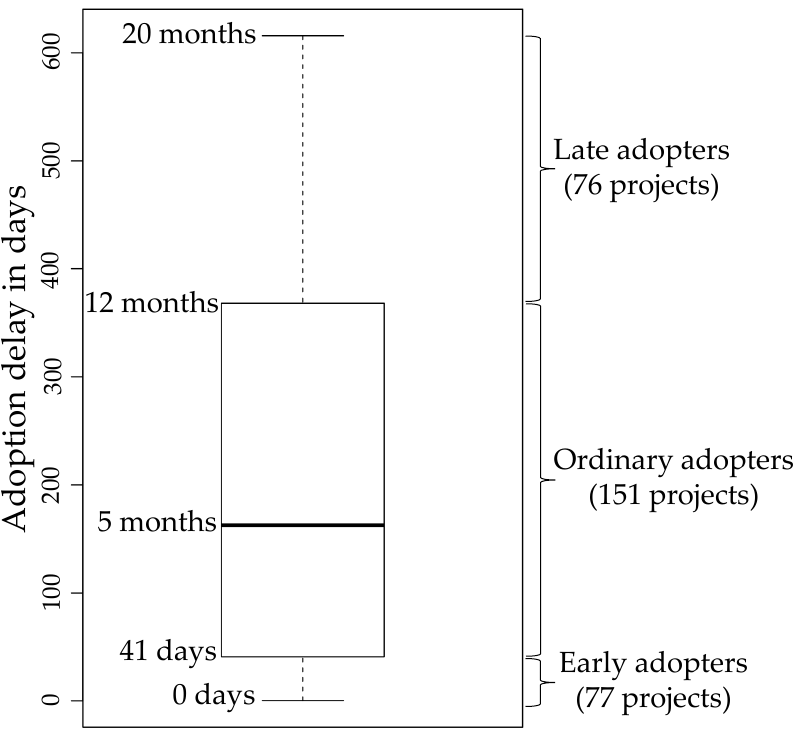}
  }
  \vspace{-5pt}
  \caption{Boxplot of the CI caching adoption delay}
  \label{fig:adoption_delay}
  \vspace{-8pt}
\end{figure}

We identify projects that adopt CI caching by analyzing their build configuration files (i.e., \texttt{.travis.yml}) and checking for the `\texttt{cache:}' directive. For each project using CI caching, we locate the first commit that introduces the `\texttt{cache:}' directive and record this as the enabling date. We then compute the number of days between this date and the announcement of CI caching for open source projects ($Dec$ $17$, $2014$). Figure~\ref{fig:adoption_delay} shows the distribution of this \textit{adoption delay}. We use the $1^{st}$ quantile (41 days) and $3^{rd}$ quantile (12.2 months) of the delay as thresholds to classify projects as \textit{early}, \textit{ordinary}, or \textit{late} adopters: (i) \textit{early} if the delay is between 0 and 41 days, (ii) \textit{ordinary} if it is between 41 days and 12.2 months, and (iii) \textit{late} if it is between 12.2 and 20 months. Based on these thresholds, we group the studied projects into five categories.

\begin{itemize}[leftmargin=18.5pt]
    \item \textbf{\textit{Non-adopters:}} projects that never adopt CI caching in their builds (899 projects).
    \item \textbf{\textit{Proactive adopters:}} projects that adopt CI caching before it was officially available (76 projects).
    \item \textbf{\textit{Early adopters:}} projects that adopt CI caching within $41$ days of being officially available (77 projects).
    \item \textbf{\textit{Ordinary adopters:}} projects that adopt CI caching within a year of being officially available (151 projects).
    \item \textbf{\textit{Late adopters:}} projects that adopt CI caching after a year of being officially available (76 projects).
\end{itemize}

To identify factors behind none, early, and late CI caching adoption, we model differences among the five project categories using project-level metrics. We employ Generalized Linear Models (GLMs) with logistic regression to analyze which factors relate to delayed or absent CI caching adoption.
We fit five GLMs (one per category) using the \texttt{glm} function from the \texttt{stats}\footnote{\url{https://www.rdocumentation.org/packages/stats}} R package. Each model compares projects in a given category (e.g., \textit{early} adopters) to all others, with a logical dependent variable: \textit{TRUE} if a project belongs to the category and \textit{FALSE} otherwise. The models use 13 project-level metrics (see Section~\ref{Metrics}) as independent variables.

\vspace{3pt}
\noindent\textbf{Seeking feedback from caching non-adopters.} 
Developers may skip CI caching simply due to unawareness or carelessness. To investigate this, we focused on projects that never use caching on {\sc Travis~CI}. Of $899$ such projects, we manually fork $160$ that (i) still had not adopted caching since data collection and (ii) had recent commits (in $2020$ or later). We trigger CI builds on these projects using the latest commit on their default branches (without further configuration changes), then exclude $71$ projects whose builds fail and four having no dependencies. For the remaining $85$ projects, we push a commit enabling dependency caching. After confirming that all builds triggered by our commits are \textit{passing}, we submit a pull request to each project asking why CI caching was not enabled.

\vspace{3pt}
\noindent\textbf{Seeking feedback from proactive and late caching adopters.} 
To understand why projects adopted CI caching \textit{proactively} or \textit{late}, we commented on commits or pull requests that enabled the CI \textit{cache} in the past. For \textit{proactive} adopters, we asked developers from $74$ projects why they enabled caching before it was available for open source projects. For \textit{late} adopters, we asked developers from $71$ projects about the reasons for their delayed adoption. We then waited several months for responses.

\subsubsection{\textbf{Findings.}}
Table~\ref{tab:caching_adoption_model} shows the results obtained from our logistic regression models of the differences between the studied projects in adopting CI caching.
In Table~\ref{tab:developer_responses}, we show a summary of the responses we obtain from developers to our submitted pull requests and comments to none/proactive/late caching adopters. We expected a quick and high acceptance rate for our pull requests, regardless of the prospective speed up CI caching might bring to CI builds, since CI caching is less likely to cause issues with CI builds. However, we observe that nearly half (48\%, 41/85) of our pull requests were approved/merged by project maintainers, with 25\% (21/85) closed, 5\% (4/85) remaining open with responses, and 22\% (19/85) still open without any response. Some pull requests took several months before being accepted or closed. Among the merged pull requests, 22\% (9/41) provided no response, while 20\% (8/41) explicitly stated they are not aware of the caching feature. A quarter (25\%, 21/85) were closed, most commonly due to projects migrating (12/21) to other CI services (e.g., {\sc GitHub~Actions}\footnote{\url{https://github.com/features/actions}})~\cite{chopra2025multici,hossain2025cigrate}.

\begin{table*}
    \renewcommand{\arraystretch}{1}
	\centering
	\caption{Modeling the differences between projects based on the caching adoption delay}
	\vspace{-8pt}
		\resizebox{.9\textwidth}{!}{
	    \begin{tabular}{llrlrlrlrlrl}
        	\toprule 
		    \addlinespace[2.5pt]

                  \multirow{2}{*}{\textbf{Project characteristic}}
                & \multirow{2}{*}{\textbf{Project-level metric}}
                & \multicolumn{2}{c}{\textbf{\textit{None/Proactive}}}
                & \multicolumn{2}{c}{\textbf{\textit{None/Early}}}
                & \multicolumn{2}{c}{\textbf{\textit{None/Ordinary}}}
                & \multicolumn{2}{c}{\textbf{\textit{None/Late}}}
                & \multicolumn{2}{c}{\textbf{\textit{Proactive/Early}}}
                \\\cline{3-4}\cline{5-6} \cline{7-8}\cline{9-10}\cline{11-12}  
                                        &                          & $\chi^2$\%	     & Signf.$^+$&   $\chi^2$\%	    & Signf.	& $\chi^2$\%	& Signf.	& $\chi^2$\%    & Signf.    & $\chi^2$\%    & Signf.\\
            \midrule    
    Language             & Ruby                       & 1.26 &  & 14.86 & * (\up) & 21.58 & * (\dn) & 0.02 &  & 15.89 & * (\up) \\[7pt]
    Code Maturity        & Size (SLOC)                & 12.8 &  & 27.13 & *** (\up) & 0.53 &  & 0.81 &  & 6.93 & \\
                         & Test density               & 0.46 &  & 0.09 &  & 21.56 & * (\up) & 46.71 & ** (\up) & 2.04  \\
                         & Dependencies               & 6.85 &  & 13.38 & * (\up) & 5.56 &  & 2.35 &  & 7.88  & \\[7pt]
    Development Activity & \# of commits per lifetime & 0.83 &  & 1.19 &  & 13.4 & . & 0.48 &  & 1.17  \\
                         & Growth rate                & 5.26 &  & 0.48 &  & 0.03 &  & 1.92 &  & 0.12 & \\
                         & \# of developers           & 12.83 &  & 6.46 & . & 0.11 &  & 17.25 &  & 25.2 & ** (\dn) \\[7pt]
    CI Activity          & CI lifespan                & 15.16 & . & 3.78 &  & 10.87 & . & 14.09 &  & 9.11 & \\
                         & Building frequency         & 10.59 &  & 5.3 &  & 1.05 &  & 4.15 &  & 7.16 & \\
                         & Configuration ratio        & 5.81 &  & 5.19 &  & 2.71 &  & 0.98 &  & 0.72 &  \\
                         & Build environments         & 9.11 &  & 5.72 &  & 0.42 &  & 2.96 &  & 22.89 & * (\up) \\
                         & Build duration             & 19.02 & * (\up) & 16.41 & ** (\up) & 22.19 & * (\up) & 8.27 &  & 0.89 & \\[2pt]
    \midrule    

                  \multirow{2}{*}{\textbf{}}
                & \multirow{2}{*}{\textbf{}}
                & \multicolumn{2}{c}{\textbf{\textit{Proactive/Ordinary}}}
                & \multicolumn{2}{c}{\textbf{\textit{Proactive/Late}}}
                & \multicolumn{2}{c}{\textbf{\textit{Early/Ordinary}}}
                & \multicolumn{2}{c}{\textbf{\textit{Early/Late}}}
                & \multicolumn{2}{c}{\textbf{\textit{Ordinary/Late}}}
                \\\cline{3-4}\cline{5-6} \cline{7-8}\cline{9-10}\cline{11-12}
                                        &                          & $\chi^2$\%	     & Signf.$^+$&   $\chi^2$\%	    & Signf.	& $\chi^2$\%	& Signf.	& $\chi^2$\%    & Signf.    & $\chi^2$\%    & Signf.\\
    \midrule    
    Language             & Ruby                       & 0.23 &  & 18.42 &  & 22.3 & * (\dn) & 5.82 &  & 23.36 & .\\[7pt]
    Code Maturity        & Size (SLOC)                & 10.08 &  & 1.19 &  & 13.9 & . & 7.25 &  & 0 & \\
                         & Test density               & 12.73 &  & 19.52 &  & 0 &  & 2.49 &  & 22.04 & .\\
                         & Dependencies               & 0.08 &  & 8.5 &  & 12.21 & . & 24.02 & * (\dn) & 14.41 & \\[7pt]
    Development Activity & \# of commits per lifetime & 1.31 &  & 0.08 &  & 3.28 &  & 0.45 &  & 3.8 & \\
                         & Growth rate                & 23.03 &  & 38.42 & * (\dn) & 2.86 &  & 2.47 &  & 5.59 & \\
                         & \# of developers           & 24.67 &  & 0.06 &  & 17.58 & * (\up) & 33.97 & ** (\up) & 14.69 & \\[7pt]
    CI Activity          & CI lifespan                & 0.1 &  & 1.58 &  & 5 &  & 0.74 &  & 1.25 & \\
                         & Building frequency         & 10.29 &  & 1.07 &  & 4.85 &  & 10.94 &  & 7.81 & \\
                         & Configuration ratio        & 0.09 &  & 0.01 &  & 2.06 &  & 0.6 &  & 0.34 & \\
                         & Build environments         & 16.44 &  & 9.01 &  & 12.1 & . & 10.12 &  & 1.38 & \\
                         & Build duration             & 0.95 &  & 2.14 &  & 3.86 &  & 1.12 &  & 5.31 & \\[2pt]
                     \bottomrule
         	 \multicolumn{5}{l}{
                 \begin{tabular}
                        {@{}l@{}} $^+$Significance codes:  0 `***' 0.001 `**' 0.01 `*' 0.05 `.' 0.1 ` ' 1
                 \end{tabular}
             }
    \end{tabular}
    }
  \label{tab:caching_adoption_model}
\end{table*}

\begin{table*}
    \renewcommand{\arraystretch}{1}
    \caption{Developer responses regarding CI caching adoption}
    \vspace{-8pt}
    \resizebox{.9\linewidth}{!}{
        \begin{tabular}{llrrL{65mm}}
            \toprule
            \textbf{Caching adoption} & \textbf{Action} & \textbf{\# (\%) responses} & \textbf{\#} & \textbf{Response}\\
            \midrule
            Non-adopters (pull requests) & \textit{Approved/Merged} & 41/85 (48\%) & 9/41 & No response given\\
                                         &                          &              & 8/41 & Not aware of caching feature\\
                                         &                          &              & 6/41 & No perceived need or benefit\\
                                         &                          &              & 5/41 & Appreciation only\\
                                         &                          &              & 4/41 & Oversight or forgot\\
                                         &                          &              & 4/41 & No particular reason or don't remember\\
                                         &                          &              & 2/41 & Assumed caching enabled by default\\
                                         &                          &              & 2/41 & Unclear or uncertain\\
                                         &                          &              & 1/41 & Technical handling issue\\\cline{2-5}
                                         
            & \textit{Open (with response)} & 4/85 (5\%) & 3/4 & No perceived benefit\\
            &                               &            & 1/4 & Introduces complexity\\\cline{2-5}
            
            & \textit{Open (no response)} & 19/85 (22\%) & 19/19 & No response\\\cline{2-5}
                                         
            & \textit{Closed} & 21/85 (25\%) & 12/21 & Migrating to other CI services\\
            &                 &              &  5/21 & No perceived benefit\\
            &                 &              &  2/21 & Unknown reason\\
            &                 &              &  1/21 & Troublesome or increases workload\\
            &                 &              &  1/21 & Not aware of caching feature\\
            \midrule
                Proactive adopters (comments) &                          &  22/74 (30\%)  &   9/22  & Caching was available unofficially (pre-release)\\
                                              &                          &                &   5/22  & Developers' misunderstanding\\
                                              &                          &                &   4/22  & Mirrored or copied from another project\\
                                              &                          &                &   3/22  & Developers cannot recall / no specific reason\\
                                              &                          &                &   2/22  & Travis~CI is no longer used\\\midrule
                
                Late adopters (comments)      &                          & 40/71  (56\%)  &  21/40  & Unaware of the caching support\\
                                              &                          &                &  12/40  & Developers are busy or have other priorities\\
                                              &                          &                &   7/40  & Developers cannot recall / no specific reason\\
         	\bottomrule
        \end{tabular}
        }
  \label{tab:developer_responses}
  \vspace{-5pt}
\end{table*}

\vspace{2pt}
\noindent\textbf{Observation 1.1. \textit{Over two-thirds of the studied projects do not adopt CI caching.}}
Despite being available to every {\sc GitHub} project, we find that CI caching has been adopted by only $30\%$ ($380$ out of $1,279$) of the projects in our dataset. 
We observe from Figure~\ref{fig:adoption_delay} that about two-thirds of the caching adopters took over a month before enabling caching in their builds. Nevertheless, $20\%$ of the caching-adopting projects adopted CI caching proactively. Based on the responses received from developers, $39\%$ of such projects were given an opportunity to use CI caching unofficially (i.e., a preview), whereas $39\%$ of the projects were configured to use caching configuration by mistake (e.g., misunderstanding or a \textit{copy/paste} from other projects).
In addition, we observe that projects that do not adopt CI caching have significantly shorter build durations, fewer developers, and fewer build configurations.
This result hints that projects that do not adopt CI caching are not as active or mature enough to seek the benefits of CI caching.

Our analysis of the responses received from projects that do not adopt CI caching reveals several reasons. The most common reason is that developers perceive no need or benefit from caching ($21\%$, 14/66), followed by projects migrating to other CI services ($18\%$, 12/66). Some developers are unaware of the caching feature ($14\%$, 9/66), and some did not respond ($14\%$, 9/66). Interestingly, some developers assumed caching was enabled by default ($3\%$, 2/66), indicating misconceptions about CI configuration, while others admitted it was an oversight or they forgot to enable it ($6\%$, 4/66). Another group had no specific reason or could not remember why caching was not enabled ($6\%$, 4/66). In addition, $22\%$ (19/85) of submitted pull requests received no response.
Still, caching support may attract developers back to {\sc Travis~CI} after prior disabling.\footnote{\href{https://github.com/ccw-ide/ccw/commit/032e253}{ccw-ide/ccw/commit/032e253}}
These results suggest CI services should communicate updates about performance-improving features like caching more effectively.

\vspace{2pt}
\noindent\textbf{Observation 1.2. \textit{Projects with shorter build durations, smaller development teams, and fewer build configurations are unlikely to adopt CI caching.}}
We observe that projects that do not adopt CI caching have significantly shorter build durations than those that do, consistent with developers stating that caching is unnecessary when builds are already fast.\footnote{\href{https://github.com/apache/storm/pull/902}{apache/storm/pull/902}}$^,$\footnote{\href{https://github.com/davidmoten/rxjava-jdbc/pull/96}{davidmoten/rxjava-jdbc/pull/96}}$^,$\footnote{\href{https://github.com/Esri/geometry-api-java/pull/283}{Esri/geometry-api-java/pull/283}}
We also observe that Non-adopting projects also have fewer developers (median = $10$) than adopting projects (median = $14$).
These findings suggest that (a) larger teams are more likely to be aware of CI caching and (b) the need for caching grows with longer build durations, as more developers are expected to wait for builds to finish.

\vspace{2pt}
\noindent\textbf{Observation 1.3. \textit{Projects with more commits and dependencies are more likely to early adopt CI caching.}}
Caching dependency binaries maximizes the benefit of CI caching. The studied projects have a median of $39$ dependencies, and our models show that dependency count is significantly associated with early CI caching adoption: each additional ten dependencies increase the odds of earlier adoption by $4\%$. Moreover, projects that do not adopt CI caching have $29\%$ fewer dependencies than those that do.
This suggests that developers of projects with few dependencies expect little build-speed improvement. For example, developers of the \texttt{test-unit/test-unit} project, which has ten dependencies, reported that CI caching brought negligible speedup.\footnote{\href{https://github.com/test-unit/test-unit/pull/123}{test-unit/test-unit/pull/123}}
It is worth noting that caching can also be applied to artifacts other than dependencies (e.g., arbitrary directories\footnote{\url{https://docs.travis-ci.com/user/caching/\#arbitrary-directories}}) that change infrequently.

\vspace{2pt}
\noindent\textbf{Observation 1.4. \textit{Late CI caching adoption is strongly associated with more build configuration activities.}}
We observe that projects that have more build configuration activity tend to adopt CI caching later.
For example, the \texttt\texttt{rubinius/rubinius} project made over $100$ build configuration changes during $5$ years of \textsc{Travis~CI} usage, but adopted caching only about $1.5$ years after it became available for open source projects.
Before adopting caching, developers made many build configuration changes, sometimes to speed up builds (e.g., by removing build jobs~\cite{ghaleb2019duration}).
Our analysis of commits and pull requests enabling caching indicates that late adoption was mainly due to (a) lack of awareness of \textsc{Travis~CI} caching support ($54\%$) or (b) higher-priority project tasks ($31\%$).

\vspace{-10pt}
\begin{rqbox}
\textbf{RQ\textsubscript{1} Summary.}
    Only $30\%$ of projects adopt CI caching. Active, mature projects adopt it early, but misconceptions and lack of awareness lead others to not adopt or to delay adoption.
\end{rqbox}

\vspace{-1pt}
\subsection{\large RQ\textsubscript{2}: How do developers maintain CI caching?}
\label{RQ2}

\subsubsection{\textbf{Motivation.}}
\textit{RQ\textsubscript{1}} shows that most projects do not adopt CI caching. Even when caching is enabled, developers need to regularly update the configuration to match code changes (e.g., adding or removing dependencies).
 Although such maintenance is important, it creates time and effort overhead similar to other build maintenance activities~\cite{mcintosh2011empirical}.
In this RQ, we investigate the historical changes to caching configurations to identify common patterns of CI cache-changing activities.

\subsubsection{\textbf{Approach.}}
We identify cache-related commits that (a) modify caching configuration (`\texttt{cache}` or `\texttt{before\_cache}`) in \texttt{.travis.yml} or (b) mention `\textit{cache}` or `\textit{caching}` in their messages. This yields $763$ commits from $380$ projects using CI caching (median: one cache-changing commit per project). After manually excluding $180$ commits unrelated to CI caching (e.g., formatting or build refactoring), we label the remaining $583$ commits by type of CI caching activity.

We generate a chronologically sorted list of cache-changing activities for each project. To identify patterns in these activities, we convert them into event logs and analyze them with the process mining framework \textsf{ProM 6}~\cite{van2005prom}. Process mining allows to uncover the underlying process of caching activities. Using \textsf{ProM 6}, we derive common maintenance patterns, event timelines, and state chart diagrams for the cache-changing events.
We then apply the \textit{Fuzzy Miner} plugin of \textsf{ProM 6} to abstract the mined process model and produce a fuzzy graph~\cite{gunther2007fuzzy}.
We call the first and last cache-changing activities of each project the \textit{first event} and \textit{last event}, respectively.
Finally, we extract the artifact types cached by the \texttt{cache} directive to determine what different projects cache in common.

\vspace{3pt}
\subsubsection{\textbf{Findings.}}
~

\noindent\textbf{Observation 2.1. \textit{Only $24\%$ of caching adopters maintain their CI caching configuration.}}
Our analysis of cache-changing commits shows that most projects (76\%) enable CI caching once and never update it,
 confirming prior work~\cite{beller2016analyzing} in which over 80\% of configuration files remain unchanged after creation.
We also find that projects with a one-time caching configuration typically cache all dependencies managed by tools such as \textit{Bundler}, \textit{Maven}, or \textit{Gradle}. Although this lets builds install only newly added dependencies, it can cause over-caching.\footnote{\url{https://eng.localytics.com/best-practices-and-common-mistakes-with-travis-ci}} As a result, dependencies no longer used by a project (e.g., outdated dependency versions) may remain stale in the CI cache.
If a CI cache is unused for some time, \textsc{Travis~CI} clears it automatically.\footnote{\url{https://docs.travis-ci.com/user/caching/\#caches-expiration}}
However, caching frequently updated dependencies can delay cache expiration and cause build problems.\footnote{\href{https://github.com/travis-ci/travis-ci/issues/3758}{travis-ci/travis-ci/issues/3758}}
\textsc{Travis~CI} allows developers to clear caches manually, but we find no evidence that they do so, as cache storage activities are not logged.
Thus, developers should avoid over-caching and consider clearing CI caches regularly.

\vspace{2pt}
\noindent\textbf{Observation 2.2. }\textbf{\textit{Developers perform five distinct maintenance activities when updating CI caching configurations.} }Table~\ref{tab:cache_changing_activities} lists cache-related activities that developers perform (at the commit level). Among the $24\%$ of projects that maintain CI caching, we observe a median of three caching-related activities per project. After initially enabling CI caching, the most common activity is updating (adding/removing) cached directories.
While updating cached directories can improve build performance, directories previously stored in the cache are not removed automatically. Moreover, only $14\%$ of cache-maintaining projects exclude unnecessary artifacts from the cache.
We also find that $50\%$ of cache-changing activities are submitted through pull requests or in response to reported issues, suggesting that (a) developers maintain cache configurations on separate branches before merging into the default branch, or (b) external contributors propose cache changes.
Developers should therefore update cached directories carefully and exclude frequently changing artifacts before uploading caches to the CI server.

\begin{table}[ht]
    \vspace{-3pt}
    \renewcommand{\arraystretch}{0.95}
	\caption{Statistics of maintenance activities of CI caching}
    \vspace{-8pt}
        \resizebox{\linewidth}{!}{
	    \begin{tabular}{L{47.7mm}rrr}
        	\toprule
        	      \textbf{Maintenance Activity} 
        	    & \textbf{Freq.}
                & \textbf{First event}
                & \textbf{Last event}\\
        	\midrule                                  
                Enable cache                             & 68.2\%              & 99.2\%                             & 80.8\%  \\
                Update cached directories                & 15.0\%              & 0\%                                & 7.9\%   \\      
                Disable cache                            & 7.4\%               & 0.8\%                              & 5.3\%   \\
                Update cache configuration               & 5.6\%               & 0\%                                & 3.2\%   \\     
                Exclude artifacts from cache             & 2.9\%               & 0\%                                & 2.1\%   \\         
                Update directories \& Exclude artifacts  & 0.9\%               & 0\%                                & 0.8\%   \\                 
         	\bottomrule
        \end{tabular}
        }
  \label{tab:cache_changing_activities}
  \vspace{-6pt}
\end{table}

\begin{figure}
  \centering
  \resizebox{0.9\linewidth}{!}{
  \includegraphics{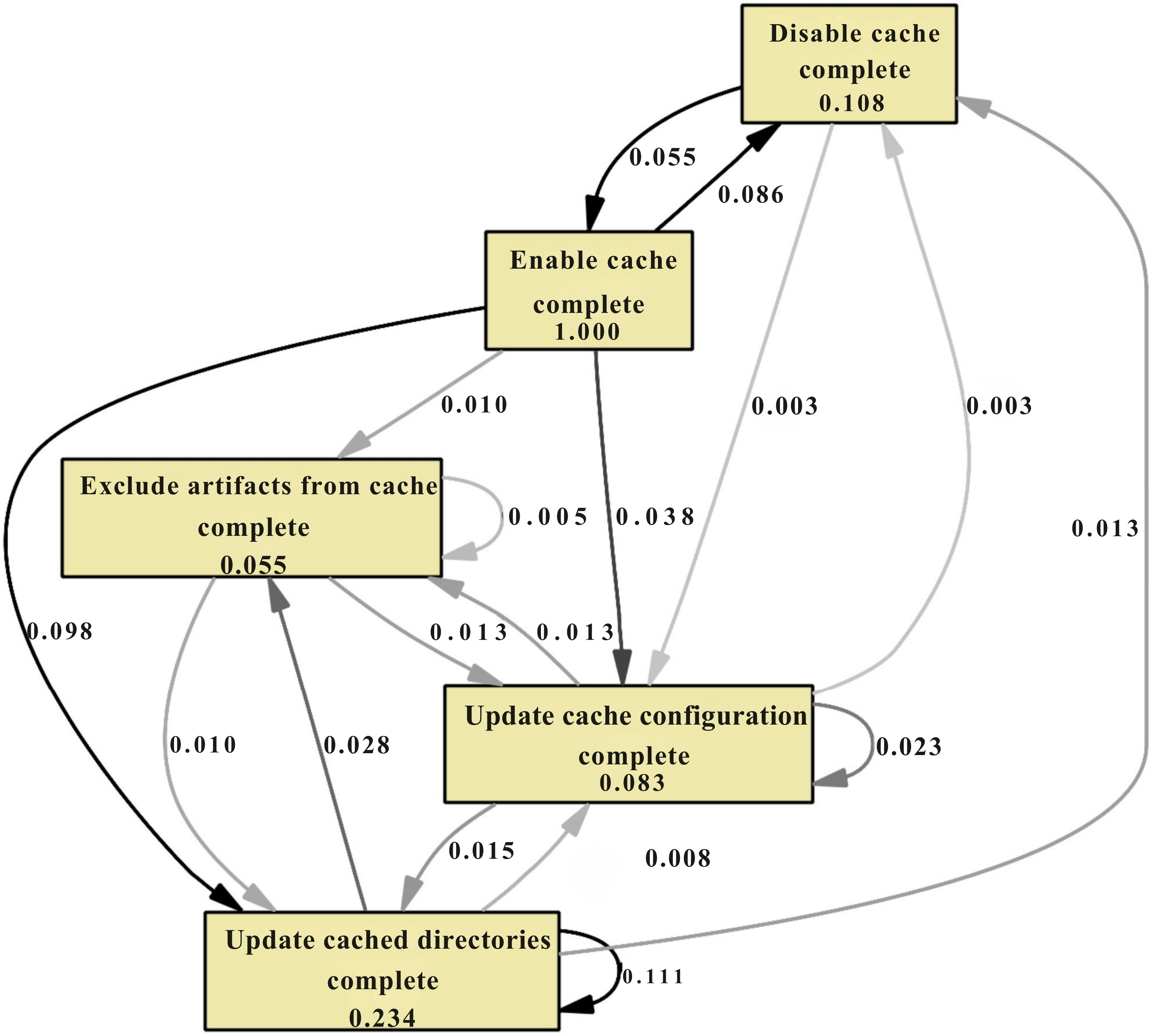}
  }
  \vspace{-5pt}
  \caption{A Fuzzy Graph of cache-changing activities}
  \label{fig:events_model}
  \vspace{-9pt}
\end{figure}

\vspace{3pt}
\noindent\textbf{Observation 2.3. \textit{The process of changing the CI cache configuration is unsystematic and repetitive.}}
Figure~\ref{fig:events_model} shows a fuzzy graph of cache-changing activities in the studied projects. Edge thickness and numeric labels indicate the percentage of events that transition from one activity to another.
We observe that cache-related events are often circular (e.g., enable-disable-enable) or repetitive (e.g., occurring multiple times in sequence).
As Figure~\ref{fig:events_model} depicts, developers frequently disable CI caching and then re-enable it in 85\% of the cases, suggesting a trial-and-error process. For example, we find that developers may disable caching temporarily until a build failure is resolved.\footnote{\href{https://github.com/ms-ati/docile/commit/8c4b9a8}{ms-ati/docile/commit/8c4b9a8}}
In addition, we observe that developers update the cached directories repeatedly (up to eight times per project), suggesting difficulty deciding what to cache. For example, the cached directories of the `SonarSource/Sonarqube' project\footnote{\url{https://github.com/SonarSource/sonarqube}} were updated eight times, adding artifacts (e.g., \texttt{node\_modules} and \texttt{.sonar/cache}) and later removing them.
About half of the caching changes occur after a median of $31$ days since the previous change, and cache updates appear irregularly throughout the project lifetime (see the timeline of cache-changing events in our replication package).
These findings suggest that (i) developers should closely monitor changes affecting the CI cache, (ii) researchers and tool builders should create tools that track historical code changes to identify frequently changing artifacts  (and should be excluded) or infrequently (and should be cached), and (iii) \textsc{Travis~CI} should use CI logs to recommend candidate artifacts for caching.

\vspace{2pt}
\noindent\textbf{Observation 2.4. \textit{The majority (67\%) of the projects cache all software dependencies.}} The {\sc Travis~CI} documentation provides simple caching examples for various languages,\footnote{\url{https://config.travis-ci.com/ref/job/cache}} which developers may interpret as best practices.
Our analysis of the \texttt{cache} directive shows that $80\%$ of Ruby builds and $45\%$ of Java builds cache all dependencies.
Other projects either customize caching from the start or initially cache an entire directory and later restrict it to subdirectories. For example, the \texttt{cloudfoundry/uaa}\footnote{\href{https://github.com/cloudfoundry/uaa/commit/53249a7}{cloudfoundry/uaa/commit/53249a7}} project first cached all artifacts in \texttt{\$HOME/.m2} (matching the documentation example), then limited caching to \texttt{\$HOME/.m2/repository}.
Dependency managers such as \textit{Bundler} and \textit{Maven} generate log files during compilation or installation. Because these files change in almost every build, they can trigger uploads of the entire cache to the CI server~\cite{travis_caching}.
We observe that only a negligible number ($4\%$) of the studied projects pre-process artifacts (e.g., exclude directories/files) before uploading the cache.
For example, the \texttt{formtastic/formtastic}\footnote{\href{https://github.com/formtastic/formtastic/commit/2c13f86}{formtastic/formtastic/commit/2c13f86}} project initially cached all artifacts in \texttt{vendor/bundle}, but later excluded generated logs (i.e., \texttt{gem\_make.out, mkmf.log}) from the cache.
We therefore encourage developers to carefully identify and exclude frequently changing artifacts from the CI cache.

\vspace{-10pt}
\begin{rqbox}
\textbf{RQ\textsubscript{2} Summary.}
    The CI cache should be maintained to handle software changes over time. However, only 24\% of projects regularly maintain their CI cache. Developers should regularly update cached artifacts to continuously benefit from CI caching.
\end{rqbox}

\vspace{-5pt}
\subsection{\large RQ\textsubscript{3}: To what extent does CI caching reduce the build duration?}
\label{RQ3}

\subsubsection{\textbf{Motivation.}}
The number and type of cached artifacts can determine how much CI caching speeds up builds. However, prior work has not examined whether CI caching actually reduces build duration in practice. In this RQ, we investigate whether current CI caching practices are associated with shorter build durations.

\subsubsection{\textbf{Approach.}}
We use Regression Discontinuity Design (RDD) \cite{cook1979quasi}, a quasi-experimental pretest-posttest design, to model changes in build duration before and after adopting CI caching. In RDD, an \textit{intervention} is assigned based on a \textit{cutoff}, allowing estimation of causal effects~\cite{imbens2008regression}. RDD assumes the pre-intervention trend would continue unchanged without the intervention. Here, the intervention is CI caching adoption, and the cutoff is the time at which caching is enabled.
We apply RDD to assess whether CI caching adoption is associated with build duration. For each project, we split builds into pre- and post-adoption and include all the builds before and after the CI adoption.
We use the \texttt{RDestimate} function from the \texttt{rdd} \textsf{R} package.\footnote{\url{https://search.r-project.org/CRAN/refmans/rdd/html/RDestimate.html}} Projects where \texttt{RDestimate} fails (e.g., due to insufficient data around adoption) are excluded. Overall, it successfully models the impact of CI caching on build duration in $75\%$ of projects.
We consider CI caching to affect build duration when the RDD estimate is significant ($p$-$value$ $\leq$ $0.05$). Each RDD model yields a Local Average Treatment Effect (LATE), the average difference between expected build durations before and after adoption. We use LATE to determine whether CI caching is associated with longer (positive LATE) or shorter (negative LATE) build durations. For instance, a negative LATE for a project indicates that CI caching is associated with reduced build duration.

\vspace{1pt}
\subsubsection{\textbf{Findings.}}

\noindent\textbf{Observation 3.1. \textit{Only a third of the projects significantly reduce build duration after adopting CI caching.}} Our project-level RDD models show that build duration becomes significantly shorter following CI caching adoption in only $33\%$ of projects, while it becomes significantly longer in $17\%$ and shows no significant change in the remaining $50\%$. To explain this, we analyze projects with either a significant increase or no significant change in build duration after adopting caching.
We observe that in $30$–$60\%$ of projects where CI caching is not associated with shorter builds, there are significant increases in (a) source lines of code, (b) build jobs, and (c) test density (details are in our replication package).
Thus, beyond CI caching, developers should consider other alternatives to speed up builds (e.g., parallelizing test suites~\cite{speeding} or removing duplicate tests~\cite{ghaleb2019duration}). These concurrent increases suggest that caching alone cannot compensate for growing build complexity, which explains why only a third of projects benefit from it.

\vspace{2pt}
\noindent\textbf{Observation 3.2.} \textbf{\textit{One-time caching configuration is strongly associated with decreased or insignificant changes in build durations.}}
In $80\%$ of projects with insignificant changes or significant increases in build duration, we observe that developers configured CI caching only once (i.e., without further maintenance).
CI build log analysis shows that the \textit{script} phase consistently dominates the overall build duration.
Thus, even when caching works as intended, it is unlikely to help much with those projects that do not heavily use external dependencies.\footnote{\href{https://github.com/FasterXML/jackson-core/pull/682}{FasterXML/jackson-core/pull/682}}
Overall, CI caching should not be treated as a ``one-size-fits-all'' solution.

\vspace{-10pt}
\begin{rqbox}
\textbf{RQ\textsubscript{3} Summary.}
    CI caching is not a “\textit{one size fits all}” solution, as two-thirds of projects see little or negative impact on build duration. Developers should regularly maintain CI caches to preserve benefits and prevent issues.
\end{rqbox}

\vspace{-6pt}
\subsection{\large RQ\textsubscript{4}: How much overhead does CI caching introduce to builds?}
\label{RQ4}

\vspace{-2pt}
\subsubsection{\textbf{Motivation.}}
Caching can improve CI build performance, but processing the cache (uploading, downloading, and storing) also adds overhead. Understanding this overhead helps developers decide whether caching is worthwhile. In this RQ, we analyze how CI builds process caches.

\vspace{-2pt}
\subsubsection{\textbf{Approach.}}
We automatically analyze $611,501$ logs of CI build jobs in which caching is enabled. We identify dependencies downloaded from the CI cache (marked `\texttt{using}') and those missing from the cache (marked `\texttt{installing}') during dependency installation. We measure the time spent for downloading and uploading the cache and installing uncached dependencies. We also detect cached but unused dependencies by comparing the dependencies downloaded from the CI cache in build $x_{n}$ with those used in the subsequent build $x_{n+1}$.
We compute cache miss rates as the percentage of cached dependencies that are not used by future builds.

\vspace{5pt}
\subsubsection{\textbf{Findings.}}
~

\noindent\textbf{Observation 4.1. \textit{CI builds perform cache uploads very frequently.}}
Our analysis of CI logs shows that $97\%$ of builds upload a cache, indicating that caches often include artifacts that change almost every run.
Uploading the cache takes a median of $12$ seconds (six times slower than downloading), while installing new dependencies takes $8.5$ seconds ($3.25\times$ slower than downloading).
Longer cache-upload times increase the risk of caching unnecessary files, such as build logs.\footnote{\href{https://github.com/twitter-archive/commons/commit/146de30}{twitter-archive/commons/commit/146de30}}
Thus, caching can add overhead if misused.
Therefore, CI services should provide cache-aware tooling~\cite{barnes2026logsieve} to help developers detect abnormal cache uploads from verbose CI logs and configure their builds to exclude unnecessary files before uploading the CI cache.

\vspace{2pt}
\noindent\textbf{Observation 4.2. \textit{One-third of the projects have a cache miss rate of $33\%$.}}
Our analysis of CI logs reveals that two-thirds of the projects have a zero cache miss rate. However, a fully hit CI cache may indicate outdated dependencies, making these projects more vulnerable to security issues~\cite{pashchenko2018vulnerable}. Kula~\etal~\cite{kula2018developers} similarly report that over $80\%$ of dependency-heavy projects do not update their dependencies.
In contrast, one-third of the projects in our dataset frequently update dependencies, causing new or updated dependencies to be installed in most builds. While updating dependencies is good practice, developers should carefully identify the most frequently updated dependencies and exclude them from the CI cache. Otherwise, builds may be delayed because (a) older dependency versions are repeatedly loaded from the cache, and (b) newer versions are installed and then re-uploaded to the cache.

\vspace{2pt}
\noindent\textbf{Observation 4.3. \textit{$27\%$ of the studied projects have unused dependencies in the CI cache}}. Caching frequently updated dependencies can leave outdated versions in the CI cache. In $73\%$ of the projects, all cached dependencies are used in builds, suggesting these projects either cache only stable dependencies or manually remove unused ones. However, in $27\%$ of the projects, dependencies used by earlier builds are not used later, implying that cache size can grow continuously if unused artifacts are not removed.\footnote{\href{https://github.com/Cockatrice/Cockatrice/issues/3781}{Cockatrice/Cockatrice/issues/3781}} In fact, developers of two projects in our dataset cite stale cached dependencies as a reason for not adopting CI caching.\footnote{\href{https://github.com/FasterXML/jackson-core/pull/682}{FasterXML/jackson-core/pull/682}}$^,$\footnote{\href{https://github.com/airlift/slice/pull/139}{airlift/slice/pull/139}} Beyond unnecessary delay, unused cached dependencies can cause storage issues, especially since a repository may keep separate CI caches for each branch and language/compiler version.

\vspace{-4pt}
\begin{rqbox}
\textbf{RQ\textsubscript{4} Summary.}
    CI caches are uploaded in $97\%$ of builds, and $27\%$ include unused artifacts. Developers should carefully choose what to cache and regularly remove stale cached dependencies.
\end{rqbox}

\vspace{-10pt}
\subsection{\large RQ\textsubscript{5}: What issues do developers encounter with CI caching?}
\label{RQ5}

\vspace{-2pt}
\subsubsection{\textbf{Motivation.}}
CI caching requires regular maintenance, yet prior work has largely overlooked the related issues developers face. Understanding these problems and their resolutions helps developers better leverage caching. In this RQ, we investigate issues that arise when projects adopt CI caching.

\vspace{-3pt}
\subsubsection{\textbf{Approach.}}
We crawl {\sc GitHub} for caching-related issues reported to the {\sc Travis~CI} team on or before $Aug$ $31$, $2016$ (the last build-triggering date in our dataset).
Using `\textit{cache}' and `\textit{caching}' as keywords, we collect $373$ caching-related issues\footnote{\href{https://github.com/travis-ci/travis-ci/issues}{travis-ci/travis-ci/issues}} from the {\sc Travis~CI} repository. These issues reflect common problems projects face when adopting {\sc Travis~CI} caching. When developers encounter a caching problem, they may comment on an existing issue instead of opening a new one.\footnote{\href{https://github.com/travis-ci/travis-ci/issues/4393}{travis-ci/travis-ci/issues/4393}}
Two co-authors manually analyze the {\sc GitHub} issues. Using open coding~\cite{corbin1990grounded}, we identify common caching issues. To resolve coding disagreements, we apply negotiated agreement~\cite{campbell2013coding}: the two co-authors collaborate on coding a sample of $50$ issues, group the resulting codes into categories, then code another $50$ issues to check for new codes, repeating this process until no new codes are identified.

\vspace{2pt}
\subsubsection{\textbf{Findings.}}
~

\noindent\textbf{Observation 5.1. \textit{Outdated or corrupted caches are the most common CI caching issues.}} Cache upload/download errors rarely cause build failures, but developers frequently report failures when CI caching is enabled.
Our analysis of cache-related issues reported to {\sc Travis~CI} shows that most failures are due to (i) corrupted caches,\footnote{\href{https://github.com/travis-ci/travis-ci/issues/6396}{travis-ci/travis-ci/issues/6396}} (ii) caches not updating dependency versions,\footnote{\href{https://github.com/travis-ci/travis-ci/issues/3758}{travis-ci/travis-ci/issues/3758}} or (iii) multiple jobs or branches overwriting a shared cache.\footnote{\href{https://github.com/travis-ci/travis-ci/issues/4657}{travis-ci/travis-ci/issues/4657}}
The {\sc Travis~CI} team replies with clarification questions or requests for supporting materials, and many caching problems are fixed in later releases.\footnote{\href{https://github.com/travis-ci/travis-ci/issues/4393\#issuecomment-219776423}{travis-ci/travis-ci/issues/4393\#issuecomment-219776423}}
When the root cause is unclear, they often suggest manually clearing the cache on the CI server,\footnote{\href{https://github.com/travis-ci/travis-ci/issues/3758\#issuecomment-106989375}{travis-ci/travis-ci/issues/3758\#issuecomment-106989375}}
 which resolves most ($62\%$) of the analyzed issues based on developer feedback. 
Details about common cache issues and the corresponding responses from the \textsc{Travis~CI} team are available in our replication package~\cite{our_replication_package}.

\vspace{2pt}
\noindent\textbf{Observation 5.2. \textit{Caching is a continuously evolving CI feature.}} Our manual analysis of CI caching issues reveals that caching keeps evolving, as the {\sc Travis~CI} team repeatedly updates the feature in response to developer requests.
For example, developers often request caching support for specific environments (e.g., \texttt{macOS}\footnote{\href{https://github.com/travis-ci/travis-ci/issues/4869}{travis-ci/travis-ci/issues/4869}}) or a simple directive to cache specific artifacts (e.g., \texttt{ccache}) instead of manually listing directories. The {\sc Travis~CI} team frequently agrees (e.g., promising support in future releases).\footnote{\href{https://github.com/travis-ci/travis-ci/issues/4997\#issuecomment-216615467}{travis-ci/travis-ci/issues/4997\#issuecomment-216615467}} In other cases, team members propose workarounds\footnote{\href{https://github.com/travis-ci/travis-ci/issues/7456\#issuecomment-296505058}{travis-ci/travis-ci/issues/7456\#issuecomment-296505058}} or disagree with the reported concerns.\footnote{\href{https://github.com/travis-ci/travis-ci/issues/4191}{travis-ci/travis-ci/issues/4191}}
Thus, developers should stay current with CI caching capabilities, which may already address previously unsolved issues.

\vspace{-10pt}
\begin{rqbox}
\textbf{RQ\textsubscript{5} Summary.}
    Developers of projects that adopt CI caching often report cache corruption (e.g., from incompatible dependency versions) and request better support for specific caching directives (e.g., \texttt{ccache}). Caching capabilities are improving.
\end{rqbox}

\vspace{-4pt}
\section{Implications}
\label{Discussion}

\vspace{-1pt}
\noindent\textbf{\large For project maintainers:}

\begin{itemize}[leftmargin=8pt]
    \item[] \textbf{What else to cache?}  
    As RQ\textsubscript{2} shows, caching is mostly limited to dependencies. However, other artifacts can benefit from caching. For example, test-generated static data can significantly increase build duration~\cite{beller2017oops,ghaleb2019duration}, so developers could cache such artifacts (e.g., test input data~\cite{korel1990automated} or search results\footnote{\href{https://github.com/thredded/thredded/commit/2f04819}{thredded/thredded/commit/2f04819}}). For Java projects, incremental compilation could cache previously compiled class files to avoid unnecessary recompilation. While {\sc Travis~CI} supports this for C/C++ via \texttt{ccache}, we found no equivalent for Java.

    \vspace{2pt}
    \noindent\textbf{No pain, no gain!}  
    Our RQ\textsubscript{2} process mining of CI cache-changing commits reveals that developers must actively maintain caches as projects evolve. In $80\%$ of projects that did not achieve significant build-time reductions, caches were not maintained, indicating that unmanaged caching is largely ineffective. While maintenance requires effort~\cite{mcintosh2011empirical}, continuously monitoring CI practices is important to sustain their benefits~\cite{santos2025need}. Our results suggest that even minimal monitoring helps preserve caching benefits.
\end{itemize}

\vspace{4pt}
\noindent\textbf{\large For CI services and tooling:}

\vspace{-2pt}
\begin{itemize}[leftmargin=8pt]
    \item[] \textbf{One caching configuration may not fit all.}
    RQ\textsubscript{3} shows that caching large files often fails to reduce build duration, and developers struggle to identify suitable directories. CI services should help (a) detect artifacts worth caching~\cite{gallaba2020accelerating} and (b) identify stale cached artifacts so developers can better tune caching strategies. Developers should also periodically update cached dependencies to prevent dependency staleness and security risks~\cite{kula2018developers}.

    \vspace{2pt}
    \noindent\textbf{Semantic linters for inefficient caching.}  
    As RQ\textsubscript{4} reports, developers tend to frequently cache directories that change often or provide minimal speed-up (e.g., \textit{JDK} packages). CI services should offer semantic linters that flag inefficient caching, beyond simple syntax checks~\cite{vassallo2020configuration,zhang2022buildsonic,bouzenia2024resource}.
    
    \vspace{2pt}
    \noindent\textbf{Simplifying selective caching.}  
    As per RQ\textsubscript{5}, excluding sub-directories is complex and error-prone: developers must either list all other sub-directories in the \texttt{cache} directive or use \texttt{before\_cache} to remove unwanted ones. CI services should support more flexible approaches (e.g., wildcards or regex) to reduce configuration effort and errors.
    
    \vspace{2pt}
    \noindent\textbf{Practical, context-aware documentation.}  
    RQ\textsubscript{2} shows that two-thirds of projects follow sample caching configurations (e.g., Ruby's \textit{bundler}), often caching all dependencies and limiting benefits. CI services could leverage historical build data to automatically suggest caching strategies based on documentation~\cite{ghaleb2025llm4ci}, such as selectively caching infrequently changing dependencies.
    
    \vspace{2pt}
    \noindent\textbf{Raising awareness of CI caching.}  
    RQ\textsubscript{1} reveals that $70\%$ of projects do not adopt caching, mainly due to ignorance or mistakenly assuming it is enabled by default. CI services should (a) recommend caching at adoption time and (b) provide estimated time savings based on build history. With nearly half of our caching-enabling PRs accepted, CI services could further promote adoption (e.g., using bots or AI agents~\cite{ghaleb2026agentscicd}) to automatically detect eligible projects and enable caching.
\end{itemize}

\vspace{0.4pt}
\noindent While our findings are based on {\sc Travis~CI}, other CI services like {\sc GitHub~Actions} share similar principles (e.g., manual caching setup)~\cite{github_actions_migration}, thus our implications likely extend beyond {\sc Travis~CI}.

\vspace{-2pt}
\section{Threats to Validity}
\label{Threats_to_Validity}

\vspace{-2pt}
\noindent\textbf{Construct Validity}
We rely on data from {\sc TravisTorrent} and cloned {\sc Git} repositories. Errors in computing metrics could affect results, but we carefully filtered and tested the data to reduce them.  
To identify CI caching issues, we used key terms likely to appear in relevant issues. This may retrieve unrelated issues; to address this, two co-authors collaboratively performed open coding to exclude them.  
Our findings come from exploratory models and manual analyses of real project data (data and scripts are publicly available~\cite{our_replication_package}). Future work could survey developers to better understand their reasons for adopting or maintaining CI caching.

\vspace{1.5pt}
\noindent\textbf{Internal Validity}
We use project-level metrics to analyze factors associated with (a) CI caching adoption delay and (b) build-duration changes, yet these metrics may not be exhaustive enough. For correlated metrics, the choice of which to retain may affect the models; to ensure reproducibility, all such choices are explicitly documented. Moreover, confounding changes coinciding with caching adoption (e.g., code growth or team changes) may affect our RQ\textsubscript{3} estimates, though we partially account for this by analyzing concurrent changes in code size, build jobs, and test density.

\vspace{1.5pt}
\noindent\textbf{External Validity}
\vspace{-1pt}
Our study analyzes builds from $1,279$ {\sc GitHub} projects on {\sc Travis~CI}, spanning up to six years and including many builds. Though the build data is from 2016, we confirm in 2021 that all projects remain active and maintained, and we successfully engage developers from 78\% of contacted projects.  
Yet, results may not generalize to other CI services, such as {\sc CircleCI} or {\sc GitHub~Actions}, which use different caching mechanisms and configuration rules that can cause different caching behaviors, misuses, or build issues. Future work should replicate and extend our study on modern CI services such as {\sc GitHub~Actions} and {\sc CircleCI}, where caching mechanisms, default behaviors, and developer practices may differ substantially from {\sc Travis~CI} as of 2016.

\vspace{-2pt}
\section{Related Work}\label{Related_Work}

\vspace{-2pt}
\noindent\textbf{Studies on CI build durations.}
    Prior work has extensively examined CI build durations. Rogers~\cite{rogers2004scaling} showed that long builds disrupt development. Although ten minutes is often considered acceptable~\cite{brooks2008team}, only $16\%$ of CI builds meet this threshold~\cite{wagner2019continuous,ghaleb2019duration}, and long executions frequently result in timeouts~\cite{weeraddana2024characterizing}. To mitigate long builds, researchers proposed less frequent integration~\cite{rogers2004scaling}, decomposing large builds~\cite{ammons2006grexmk}, and removing unnecessary dependencies or tests~\cite{hilton2017trade}. CI caching is a promising way to reduce build time~\cite{ghaleb2019duration,ghaleb2022interplay}, and caching images and dependencies can further accelerate builds~\cite{gallaba2020accelerating}, while developers also apply custom acceleration strategies beyond caching~\cite{yin2024developer}. Soares~\etal~\cite{soares2022effects} reviewed CI's effects, and Zheng~\etal~\cite{zheng2025github} analyzed CI workflow failures.
    Yet, the actual impact of caching on CI builds remains unclear and is likely context-dependent. Our study examines current CI caching practices and their associated performance improvements or overhead.
    
\vspace{1.5pt}
\noindent\textbf{Studies on CI configuration.}
    Several studies have explored CI configuration practices. Hilton~\etal~\cite{hilton2017trade} reported that configuring CI builds is challenging, and Widder~\etal~\cite{widder2019conceptual} found that developers may abandon CI due to long builds, hard-to-diagnose failures~\cite{ghaleb2019noise}, and configuration complexity~\cite{ghaleb2025android,abrokwah2025complexity}. Vassallo~\etal~\cite{vassallo2020configuration} identified smells in CI configurations. Recent work has analyzed CI configuration and testing practices in mobile apps~\cite{ghaleb2025android,zhou2026roleci,parsazadeh2026instrumentation}, including workflow failures~\cite{zheng2025github}, automation and reuse~\cite{delicheh2026automation}, and the use of multiple CI services~\cite{rostami2023usage,chopra2025multici}. Studies on {\sc GitHub~Actions} have examined CI workflow maintenance~\cite{valenzuela2024hidden}, complexity and compliance~\cite{abrokwah2025complexity}, developers' perceptions~\cite{saroar2023developers}, security issues~\cite{koishybayev2022characterizing}, and outdated actions~\cite{decan2023outdatedness}.
    Despite this, CI caching practices remain understudied. We address this gap by analyzing their adoption, maintenance, and impact on build performance.

\vspace{1.5pt}
\noindent\textbf{Studies on CI build issues.}
    Prior work examined CI build issues. Hilton~\etal~\cite{hilton2017trade} studied assurance, security, and flexibility trade-offs, and Pinto~\etal~\cite{pinto2018work} analyzed developers' uncertainties and confusion. CI issues and bad practices have also been studied using Stack Overflow data~\cite{zampetti2020empirical,ouni2023empirical}. Tools such as CI-Odor~\cite{vassallo2019automated} and BuildSonic~\cite{zhang2022buildsonic} detect CI anti-patterns and {\sc Travis~CI} configuration smells. Khatami~\etal~\cite{khatami2024catching} cataloged recurring anti-patterns in {\sc GitHub~Actions} workflows.
    However, CI caching issues remain largely underexplored. Our study investigates the challenges developers face when adopting CI caching and how to address them.
    
\vspace{-2pt}
\section{Conclusion}\label{Conclusion}
\vspace{-2.1pt}
This paper presents an empirical study of CI caching practices in {\sc Travis~CI}, analyzing $513{,}384$ builds from $1{,}279$ {\sc GitHub} projects to examine adoption, maintenance, impact on build duration, and common issues. Our results show that $70\%$ of projects do not adopt CI caching, primarily due to limited awareness and varying project characteristics, and $76\%$ of projects do not maintain their caches, highlighting that effective caching requires ongoing updates. We find that two-thirds of projects experience no significant reduction in build duration after enabling caching, indicating that benefits depend on proper maintenance. Moreover, two-thirds of projects ($67\%$) cache all dependencies, which can lead to unnecessary cache uploads when artifacts change frequently, and caching may also introduce issues such as cache corruption or limited support for certain scenarios. These findings underscore the need for researchers, tool builders, and developers to collaborate on reducing CI cache maintenance overhead and improving caching support.

\vspace{1.5pt}
\noindent\textbf{Future work.}
We plan to survey and interview developers to better understand awareness and practices related to CI caching, including the security implications of stale cached dependencies and the automated detection of optimal artifacts to cache based on project context and build history. We also plan to replicate our study on modern CI services such as {\sc GitHub~Actions} and {\sc CircleCI} to assess whether our findings generalize beyond {\sc Travis~CI}.

\vspace{-2pt}
\begin{acks}
\vspace{-2.1pt}
This work is funded by the Natural Sciences and Engineering Research Council of Canada (NSERC): RGPIN-2025-05897.
\end{acks}

\clearpage
\balance
\bibliographystyle{ACM-Reference-Format}
\bibliography{paper}

\end{document}